# Role of thermal and photo annealing on nonlinear optical response of $Ge_{30}Se_{55}Bi_{15}$ thin films

Adyasha Aparimita[1], P. Khan[2], J.R. Aswin[3], K.V Adarsh[3], and R. Naik[1,4*]

[1]*Department of Physics, Utkal University, Vanivihar, Bhubaneswar, 751004, India*

[2]*Department of Physics and Bernal Institute, University of Limerick, V94 T9PX Limerick, Ireland*

[3]*Department of Physics, Indian Institute of Science Education and Research, Bhopal, 462023, India*

[4]*Department of Engineering and Material Physics, ICT-IOC Odisha campus, Bhubaneswar,751013, India*

In this article, we employed nanosecond *Z*-scan technique to demonstrate the nonlinear optical response in $Ge_{30}Se_{55}Bi_{15}$ thin films after thermal and photo annealing. The intensity dependent open aperture *Z*-can traces reveal that for all the samples, i.e. as-prepared, thermal and photo annealed thin films exhibit reverse saturable absorption (RSA). The experimental results indicate that both thermal and photo annealing can be efficiently used to enhance the nonlinear absorption coefficient ($\beta$) compared to as-prepared sample. We further demonstrate that $\beta$ value of thermally annealed and as-prepared samples increase significantly at higher intensities. On the contrary, $\beta$ for photo annealed sample doesn't exhibit appreciable changes against the intensity variation.

- Corresponding author Email: ramakanta.naik@gmail.com

## I. INTRODUCTION

Amorphous chalcogenide glasses (ChG) stands apart from their counterpart optical materials because of their unique linear and non-linear optical properties. For example, they possess high linear refractive indices[1] (2.0-3.0 at 1.55µm) and high third order nonlinear refractive index (~1000 times that of silica)[2,3]. The later has attracted significant interest in theoretical and experimental investigations owing to their numerous applications in various fields which includes optical limiting[4], optical switching[5], multi-photon polymerization[6], photo-dissolution[7]etc. Till date, most of the previous experiments exploring the non-linear optical properties have been conducted on Ge and As-based ChGs[8-10]. For example, Huang and co-workers observed strong enhancement in third order non linearities as a result of formation of defect like bonds in Ge-Sn-Se chalcogenide glasses[11]. On the other hand, Smektala and colleagues demonstrated a strong variation of the nonlinear refractive index of $As_2S_3$ and $As_2Se_3$ glasses with the input laser intensity[12].

In spite of high photosensitivity of ChGs, their application get limited because of poor thermal stability. An efficient way to overcome such issue is doping a third impurity elements, such as Bi, Sb with more metallic character into the ChG matrix which can improve the thermal stability significantly with added functionalities[13,14]. For example, with the incorporation of metallic Bi, glass formation region gets expanded and stabilized for Ge-Se-Bi system[15]. The chemical durability and IR transmission also increases by the addition of Bi into Ge-Se system. Nevertheless, for practical applications, it is important to tune the linear and non-linear optical properties of Ge-Se-Bi system. Thermal and photo annealing are considered to be efficient techniques to alter the linear and non-linear optical constants of ChGs. Although, the effect of such annealing has been studied on the linear optical properties, the implications remain unknown for the non-linear optical response.

To gain new insights on the third order non-linear optical properties of Ge-Se-Bi ChGs upon thermal and photo annealing, we employed conventional open aperture Z-scan technique. Z-scan traces reveal that upon resonant excitation of 1064 nm, all the samples i.e. as-prepared, thermal and photo annealed samples undergo RSA. A comparative study reveals that non-linear absorption coefficient (β) increases significantly upon thermal and photo annealing.

## II. EXPERIMENTAL

The bulk $Ge_{30}Se_{55}Bi_{15}$ was prepared by using melt quench technique from the high pure elements Ge, Se and Bi (5N, Aldrich and Sigma Chemical Company). The highly pure elements were weighted in the stoichiometric ratio and placed into a pre-cleaned quartz ampoule. The ampoule was then sealed under vacuum at a pressure of about ~ $5\times10^{-4}$ Torr. The ampoules were placed in a programmable rocking furnace and slowly heated up to $950^0$ C for about 36 hours with continuous gentle rocking to ensure homogeneity. After full homogenization of the melt, the ampoules were rapidly quenched in ice cold water.

Thin films of $Ge_{30}Se_{55}Bi_{15}$ was deposited from the prepared bulk material by thermal evaporation technique on glass substrates at a base pressure of ~ $5 \times 10^{-5}$ Torr. The substrate temperature was maintained at room temperature and the deposition rate was fixed at 5 nm per second during the deposition process. The substrates were rotated slowly to get homogeneous and uniform film. We prepared films of thickness ~ 1 µm, confirmed from thickness crystal monitor. The films were thermally annealed for 2 hours at $120^0$ C. For photo annealing, the sample was irradiated with 532 nm continuous wave laser for ~ 2 hours with illumination intensity of 50 mW $cm^{-2}$.

To explore the nonlinear optical properties of the films, we employed open aperture Z-scan technique which measures the total transmittance as a function of incident laser intensity. The schematic diagram of open aperture Z-scan technique is presented in Fig.1. In this method, the thin films were illuminated with 1064 nm, 5 ns pulses of Nd:YAG laser. We used a fixed repetition rate of 10 Hz to avoid sample heating and photo-damage.

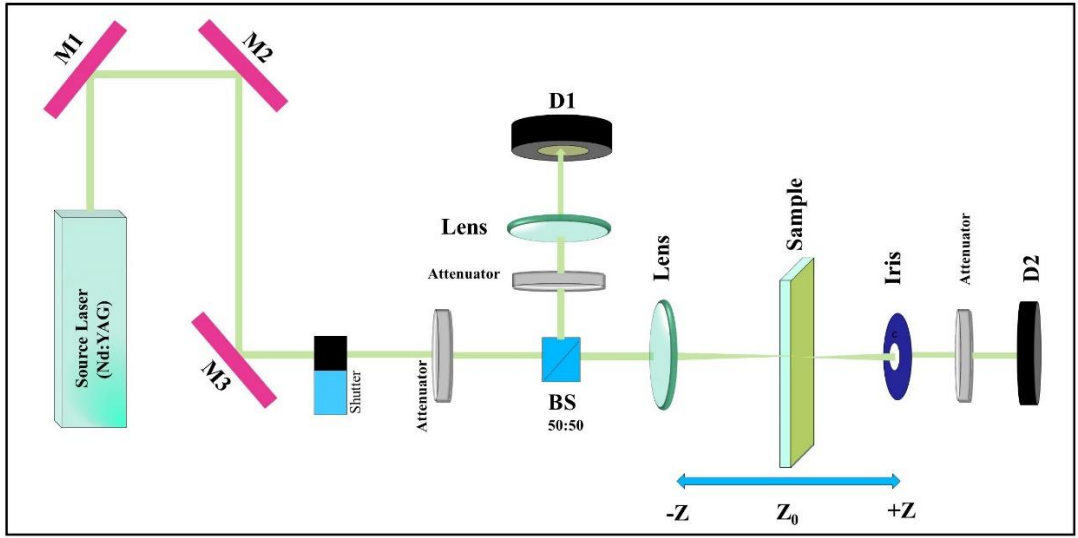

**Fig.1** Schematic diagram of open aperture Z-scan technique.

## III. RESULTS AND DISCUSSIONS

First, we investigated the effect of thermal and photo annealing on the linear optical properties of as-prepared $Ge_{30}Se_{55}Bi_{15}$ thin films by recording the optical absorption spectra of the samples as shown in Fig. 2a. The optical absorption coefficient (α) was calculated from the absorbance spectra from the following formula[16]

$$\alpha = \frac{2.303A}{t} \qquad (1)$$

where A and *t* are the absorbance and the thickness of the thin film, respectively. Then, we calculated the optical band gap of each sample by using Tauc equation[17]

$$(\alpha h\nu) = B(h\nu - Eg)^2 \qquad (2)$$

where α, h, ν, $E_g$ and B are the absorption coefficient, Plank's constant, frequency, optical band gap, and a constant (Tauc parameter) respectively. A straight-line fitting of the plot $(αhν)^{1/2}$ vs hν to the energy axis gives the value of optical band gap which is shown for all samples in Fig. 2(b).

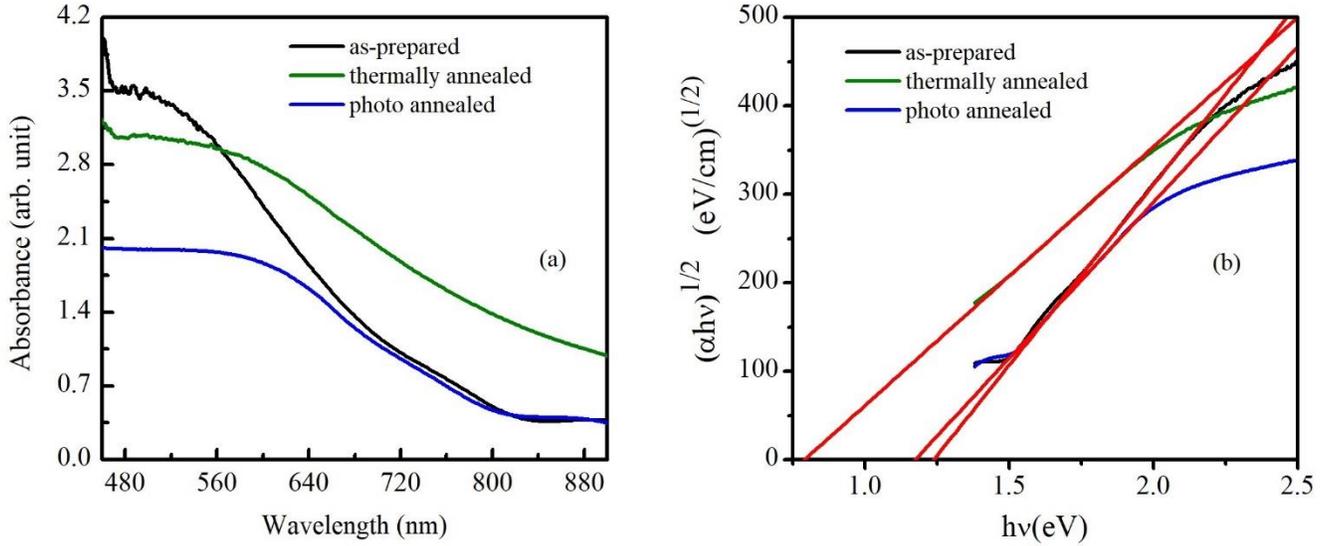

**Fig.2** (a) Optical absorbance spectra of the as-prepared, thermally and optically annealed films. **(b)** Tauc plot for the respective films for the calculation of the band gap.

From the Tauc plot, the band gaps are found to be 1.22, 0.77, and 1.17 eV for as-prepared, thermally annealed, and photo annealed $Ge_{30}Se_{55}Bi_{15}$ films, respectively. Clearly, the optical band gap decreases upon thermal and photo annealing. The decrease in the band gap upon thermal annealing can be explained by Mott and Davis model which says that width of the localized states and band gap depends on the degree of disorderness and defects present in the amorphous structure[18]. When the as-prepared sample is thermally annealed below the glass transition temperature, additional dangling bonds are formed which leads to the formation of the defect states. Consequently, width of localized states increases which eventually results in the decrease in band gap for the thermally annealed films. On the other hand, the reduction in optical band gap by photo annealing is the direct consequence of conventional photodarkening process[19,20] and the effect of which is less than the thermal annealing. We believe that the tunability of the optical band gap [21] of the sample can be useful for many optical devices.

In spite of the promising linear optical properties of ChGs as active materials in optoelectronic devices, an extremely important issue to be addressed is the non-linear optical properties which makes them a favorable candidate to fabricate several non-linear optical devices. In this regard, we present an interpretation of the third-order nonlinear ($\chi^{(3)}$) optical properties of as-prepared, thermally and optically annealed $Ge_{30}Se_{55}Bi_{15}$ chalcogenide thin films.

We employed conventional open aperture Z-scan technique to unveil the non-linear optical properties. In this context, Fig 3. (a)-(c) show the open aperture Z-scan traces at three peak intensities of resonant excitation at 1064 nm for as-prepared, thermally and photo annealed $Ge_{30}Se_{55}Bi_{15}$ thin films, respectively. At these very low intensities, for all the samples, the normalized transmittance shows a gradual decrease when the position of the sample (Z) approaches zero (focal point). The least possible value of transmittance at the focal point is represented by the dip in normalized transmittance curve. Such curves clearly illustrate that RSA takes place in our sample similar to the observation of Elim and co-workers.[22] RSA has many potential applications in optical pulse compressor, optical switching, and laser pulse narrowing etc[22]. With the increase in laser intensity, the transmittance decreases gradually that is a typical behavior of optical limiting process which plays important role in protecting optoelectronic detectors as well as human eye from intense laser sources[23].

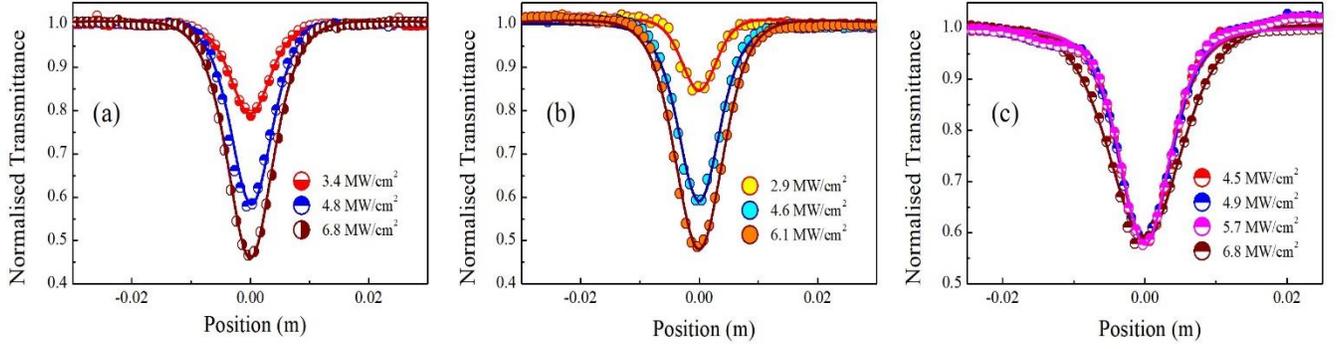

**Fig. 3.** Normalized transmittance as a function of input intensity of (a) as-prepared (b) thermally annealed (c) photo-annealed $Ge_{30}Se_{55}Bi_{15}$ thin films.

To quantify RSA process, we computed the non-linear absorption coefficient (β) and saturation intensity ($I_s$) of all the samples from the following propagation equation[24].

$$\frac{dI}{dz} = -\alpha(I)I \tag{3}$$

where α(I) is intensity dependent absorption coefficient and z is the propagation distance in the sample. Here α (I) can be written as

$$\alpha(I) = \frac{\alpha_0}{1+\frac{I}{I_s}} + \beta_{RSA}I \tag{4}$$

where $\alpha_0$ is the linear absorption coefficient, β is the RSA coefficient and $I_s$ is the saturation intensity. However normalized transmittance as a function of position z for RSA can be expressed as:

$$T_N = \frac{1}{q_0\sqrt{\pi}}\int_{-\infty}^{+\infty} \ln(1 + q_0 e^{-t^2})\, dt \tag{5}$$

Where $q_0 = \frac{\beta I_0 L_{eff}}{1+\frac{Z^2}{Z_0^2}}$ and $L_{eff} = \frac{(1-e^{-\alpha L})}{\alpha}$   Were $I_0$, $z_0$, L and $\alpha$ the peak intensity at the focus (z = 0), the Rayleigh length, sample thickness, and linear absorption coefficient, respectively[23, 25].

RSA coefficients were extracted from the best fit of normalized transmittance curve. The strength of RSA is determined by nonlinear absorption coefficient ($\beta$). To obtain an overall picture, we plotted the variation of $\beta$ and $I_{sat}$ as a function of laser intensity for three different samples in Fig. 4(a) and 4(b) respectively, and the results are summarized in Table 1. It can be seen that for both as-prepared and thermally annealed samples, $\beta$ increases with laser intensity, which is a true signature of RSA process. On the contrary, for photo annealed sample, $\beta$ doesn't exhibit any significant changes against the intensity variation. $I_{sat}$ exhibits complementary behavior to $\beta$, i.e. decreases with laser intensity for as-prepared and thermally annealed samples and likewise remains invariant for optically annealed sample.

**Table 1.** *Variation of $\beta$ and $I_{sat}$ value with different peak intensity for as-prepared, annealed and illuminated $Ge_{30}Se_{55}Bi_{15}$ thin films.*

| Sample | Intensity (MW/cm$^2$) | $\beta$ (cm/GW) | $I_{sat}$ (GW/cm$^2$) |
|---|---|---|---|
| *As-prepared* | 3.4±0.2 | 22960±530 | 16.8±0.8 |
|  | 4.8±0.1 | 34320± 980 | 10.8±0.9 |
|  | 6.8±0.3 | 37510±810 | 9.1±0.6 |
| *Annealed* | 2.9±0.1 | 61790±6110 | 25.5±3.3 |
|  | 4.6±0.2 | 85310±5100 | 21.0±2.3 |
|  | 6.1±0.1 | 94200±3430 | 16.5±1.1 |
| *Illuminated* | 4.5±0.1 | 46470±2650 | 15.9±2.2 |
|  | 4.9±0.2 | 41750±2090 | 16.9±2.0 |
|  | 5.7±0.1 | 40080±1450 | 18.6±1.6 |
|  | 6.8±0.3 | 38700±1080 | 16.6±1.1 |

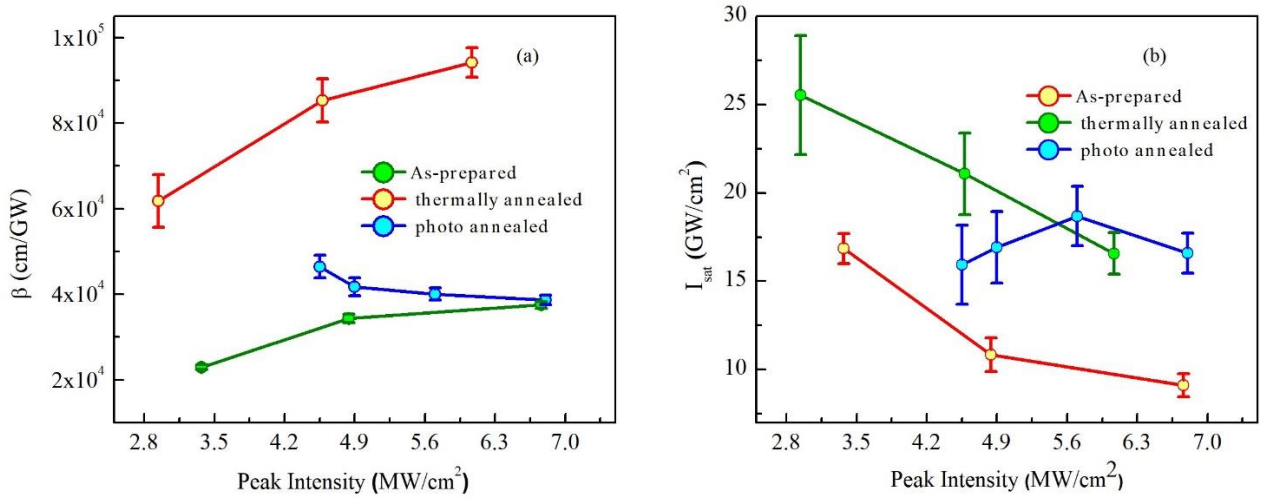

**Fig. 4**. (a) Variation of RSA coefficient β and (b) $I_{sat}$ of three samples as a function of laser peak intensity.

To quantify the effect of thermal and photo annealing on the non-linear optical properties, we compared the non-linear absorption coefficients β and $I_{sat}$ of all the samples at comparable peak intensity. From the table we learnt that β increases from (34320±980) cm/GW to (85310±5100) cm/GW after thermal annealing; whereas for photo annealed films, β increases to (46470±2650) cm/GW (table 2). Precisely, we found that β is significantly increased in thermally and optically annealed films compared to as-prepared one, however, the enhancement is stronger by two times for thermally annealed samples compared to photo annealing.

*Table 2*. *Experimentally calculated values of $E_g$, β and $I_s$*

| Samples | $E_g$ (eV) | β (cm/GW) | $I_{sat}$ (GW/cm$^2$) |
|---|---|---|---|
| *As-prepared* | 1.22 | 34320±980 | 10.8±0.9 |
| *Annealed* | 0.77 | 85310±5100 | 21.0±2.3 |
| *Illuminated* | 1.17 | 46470±2650 | 15.9±2.2 |

Thus, the result clearly indicates that the enhancement of $\beta$ could possibly be a direct consequence of band gap reduction. In literature [25,26], it is shown that non-linear response (RSA) in the semiconductor follows an inverse cubic relation with band gap. i.e.

$$\beta^{(2)} = K^{(2)} \frac{\sqrt{E_p}}{n^2 E_g^3} \times \frac{(2h\vartheta/E_g - 1)^{\frac{3}{2}}}{(2h\vartheta/E_g)^5} \tag{6}$$

where $K^{(2)}$=3.10 eV$^{5/2}$cm/MW and $E_p$ (Kane energy) are material independent constant, n is the refractive index of material and the ratio $\left(\frac{(2h\vartheta/E_g - 1)^{\frac{3}{2}}}{(2h\vartheta/E_g)^5}\right)$, depends on the photon energy and band gap, which will help to understand the band structure of compound. We found from Fig. 2(b) that band gap decreases for thermally and optically annealed samples compared to as-prepared sample which leads to the enhancement of β as predicted from the theoretical perspective.

## IV.  CONCLUSIONS

In conclusion, we demonstrated the non-linear optical properties of Ge$_{30}$Se$_{55}$Bi$_{15}$ thin films by open aperture Z-scan measurements. We investigated the role of thermal and photo annealing on the non-linear optical response of the sample. All the samples exhibit RSA as a function of laser intensity. Our results indicate that RSA coefficient (β) increases after thermal and photo annealing with stronger effect is observed for thermal annealing. By comparing the linear and non-linear optical properties, we conclude that the significant increase in β is correlated with the reduction in optical band gap. The observed non-linear optical absorbance is described on the basis of density of defect states and disorders present in the mobility gap. We believe that the remarkable enhancement of RSA holds tremendous potential applications in many optical devices like optical switching and limiting.

**ACKNOWLEDGEMENTS**

The authors thank Board of Research in Nuclear Science (BRNS) for financial support (Grant Number: 37(3)/14/02/2016-BRNS/37016) and Department of Physics, Indian Institute of Science Education and Research, Bhopal (IISER) for non-linear optical property study.